# Collaborative Driving Support System in Mobile Pervasive Environments

Nevin Vunka Jungum*, Razvi M. Doomun, Soulakshmee D. Ghurbhurrun and Sameerchand Pudaruth
Computer Science and Engineering Department
University Of Mauritius
Réduit, Mauritius
nevin.vunka@umail.uom.ac.mu, {r.doomun, s.ghurbhurrun, s.pudaruth}@uom.ac.mu

*Abstract*— **The Bluetooth protocol can be used for inter-vehicle communication equipped with Bluetooth devices. This work investigates the challenges and feasibility of developing intelligent driving system providing time-sensitive information about traffic conditions and roadside facilities. The architecture for collaborative vehicle communication system is presented using the concepts of wireless networks and Bluetooth protocol. We discuss how vehicles can form mobile ad-hoc networks and exchange data by the on-board Bluetooth sensors. The key design concepts of the intelligent driving service infrastructure are analyzed showing collaborative fusion of multiple positional data could give a better understanding of the surrounding traffic conditions for collaborative driving. The technical feasibility of using Bluetooth for data exchange among moving vehicles is evaluated.**

I. INTRODUCTION

Traffic congestion and accidents in many countries cost several milliards of dollars a year in lost productivity and wasted energy. Pervasive computing technologies are being employed in the development of intelligent transport systems or driving support systems to try to alleviate these costs. Such systems seek to bring together information and telecommunications technologies in a collaborative scheme to improve the safety, efficiency and productivity of transport networks and driving experience. Research is currently focused on filling in the knowledge and technology gaps in pervasive, mobile ad hoc wireless systems for a range of transport applications. Electronic sensor devices, such as Bluetooth nodes, could be directly integrated into the transport infrastructure, and into vehicles themselves, with the aim of better monitoring and managing the movement of vehicles within road transport systems. Computers are already incorporated into modern cars via integrated mobile phone systems, parking sensors and complex engine management systems. Hence, mobile wireless systems are beginning to be proven as a future tool that will enable the joining up of vehicles, individuals and infrastructure into a single 'connected' intelligent infrastructure system. Implanting this technology in infrastructure (such as environmental sensors in lampposts, embedded in vehicles, in goods and even connecting to individuals through their PDAs, or mobile phones) offers potential for a more 'all-seeing, all knowing' intelligent transport system infrastructure. Vehicles are capable of receiving and exchanging information 'on the move' via wireless technologies and be able to communicate with devices integrated into transport infrastructure, alerting drivers to traffic congestion, accident hotspots, and road closures. Alternative routes could be relayed to in-car computers, speeding up journey times and reducing road congestion. This would bring added coordination to the road transport system, enabling people and products to travel more securely and efficiently. Safety of road travel can be increased and traffic congestion can be minimized if vehicles can be made to form groups for communicating data among themselves. If, for example, vehicles were continually in wireless communication with the infrastructure (through small wireless sensors embedded in the infrastructure), new paradigms for traffic monitoring and control could be considered, road space allocated more efficiently and incidents dealt with in an optimum way. If vulnerable users had such wireless devices, the infrastructure could warn vehicles to slow down and drivers to be more vigilant. Indeed, wireless devices attached to children could, for example, warn drivers that children are playing out on the street, just around the corner, so they should reduce speed now.

Hence, research in communication among vehicles is being conducted under the paradigm of Inter-Vehicle Communication (IVC). In the last couple of years, IVC has emerged as a promising field of research, where advances in Wireless and Mobile Ad-Hoc Networks can be applied to real-life problems and lead to a great market potential [1]. Numerous major automobile manufacturers and research centers are investigating the development of IVC protocols and systems, and the use of inter-vehicle communication for the establishment of Vehicular Ad-Hoc NETworks (VANETs) [2], [3], [4], [5], [6]. In designing vehicular communication systems, a good design should follow certain criteria, such as intuitive and easy-to-use interfaces, reasonable range of communication, confidentiality and privacy. Bluetooth can be used to form wireless ad hoc networks, or networks based on a collection of wireless nodes that dynamically form a temporary network as long as these devices are within a sufficient range. The flexibility in ad hoc networks is what makes it a suitable choice for modeling common on-road vehicle-to-vehicle communication scenarios -where multiple cars would be traveling at different rates but potential connectivity exists. For example, a car traveling at a given rate establishes connection with an approaching car within range to form an ad-hoc network. In practical terms, a motorist would be able to communicate with adjacent vehicles, as long as those vehicles are in range of the vehicular communication system. Such a device could be built upon the application layer of Bluetooth's protocol stack, in effect not changing any internal behavior of Bluetooth, but integrating the specification into software.

The objective of the Multiple IVC concept is to demonstrate query processing in a mobile environment. In a group of IVC vehicles, the vehicles will travel on the highway, in live traffic, in a convoy with a distance range between successive vehicles. They will process queries of the type "what is the average speed one mile ahead?" Processing such a query involves wireless mobile multi-hop communication, and mobile database collaboration.

The rest of the paper is as follows. Section II presents the intelligent transport service environment with some motivating scenarios. Section III gives a detailed description of the system architecture. Section IV the simulation results. Finally, Section V concludes the main contribution of the paper and future works.

## II. INTELLIGENT TRANSPORT ENVIRONMENT

This section presents an overview of the different hardware requirements that are assumed in the collaborative driving environment. Specific scenarios that occur in the intelligent collaborative driving system are discussed and location-based access control to resources is elaborated.

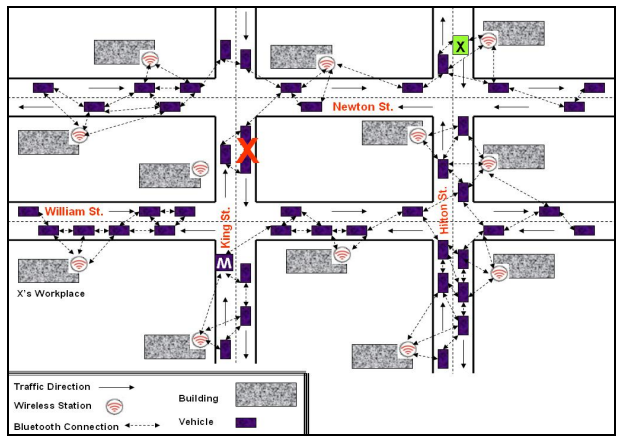

Figure 1. A collaborative scenario

### A. Vehicle and Environment Settings

Vehicles participating in the service infrastructure are equipped with an embedded microprocessor with a display interface, a GPS receiver, a class 1 Bluetooth sensor node, and an onboard diagnostics (ODI) interface. Some vehicles may have alternative wireless network connectivity support based on an on-board cellular communication device. The ODI is used to acquire a small set of data values from mechanical and electronic sensors mounted on the vehicle. All subsystems (GPS, ODI, wireless networking and Bluetooth links) are connected and forward data to the embedded microprocessor. A navigation software system enables the association of the vehicle's geographic position to an internal data-structure representing the road networks of a large geographic area around the vehicle. This type of data structure is easily constructed from publicly available geographic referencing systems [7], [8]. Alternatively, the vehicle can also be equipped with a Bluetooth-enabled mobile phone onto which the client application supports collaborative navigation system.

Vehicles in proximity range establish ad-hoc networks on-the-fly to create a service infrastructure through their bluetooth connections depicted by double-headed dashed arrows.

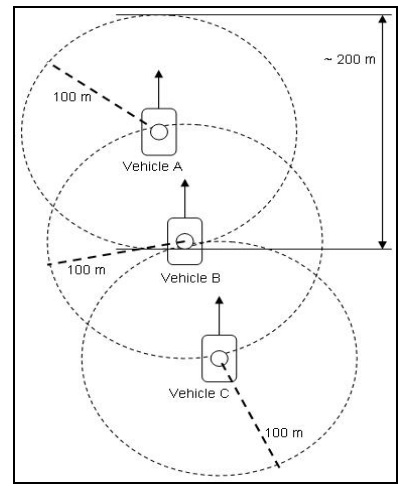

Figure 2. Vehicle Bluetooth Range

Consider Vehicle A, in figure 2, is traveling at 90 Km/h and Vehicle B is traveling at 106 km/h. The difference of 16 km/h equates to approximately 4.5 m/s. After coming into range, Vehicle B would need to travel 200m more than Vehicle A for it to exceed the Bluetooth range. Using these speed values the two vehicles would be in range for a period of approximately 44 seconds. During this time period vital information can be exchanged between the vehicles depending on the envisaged application.

### B. Motivating Scenarios

In this part, we present two motivating scenarios based on Figure 1 and 3. Collaboration among vehicles relates to the exchange of messages. Messages are in the form of (a) queries where the vehicle that makes the query expect a response, which we called *Query Messages* and (b) alerts where the vehicle that creates the alert do not expect any response, which we called *Alert Messages*.

*1) Query Messages*

Consider vehicle X (green colored) located in Hilton Street on the top-right of Figure 1. The vehicle is in fact moving to the west of William Street. The driver of vehicle X is also interested in finding the nearest drive-in coffee shop along the way. He asks the on-board navigation system for the different possible routes and their respective traffic conditions and for the location of drive-in coffee shops along those routes. Notably, one possible way to reach the west of William Street goes through Newton Street, King Street and then William Street. The other way, is to continue on Hilton Street and then take a right-turn in William Street which is only a few meters down the road from the present location of vehicle X. Therefore, the service infrastructure should try to come up with a reply to the driver's requests, before the driver decides whether or not to take the Newton Street.

The information requested by the driver of vehicle X can be *computed* out of data available on vehicles and roadside facilities located in the road segments specified by the green vehicle's query. For example, the traffic-flow on the segment of Newton Street shown in Figure 1 can be derived by estimating the average speed of vehicles moving on that segment for a short period of time; congestion in that road segment can be established from a consistently low average speed and/or a high density of vehicles on that road. Similarly, the operation of a coffee shop can be deduced from information dispatched by the coffee shop's wireless access point, which specifies the type of service offered (selling coffee), the business address, and coffee prices.

To retrieve such information, the on-board system of the green vehicle has to translate end-user queries into a sequence of location-sensitive queries. Each of these queries should be routed toward its targeted location of interest via the service infrastructure. Upon arrival to its destination area, the query must be picked up by the local service infrastructure that represents the target area. Nodes of that infrastructure (vehicles and/or roadside services) collaborates on-the-fly to compute a reply, which is dispatched back to the location where the query came from.

Messages are dispatched in two different basic ways (refer to Figure 3):

- **Vehicle**-to-**Vehicle**
- **Vehicle**-to-**Local Server**-to-**Vehicle**

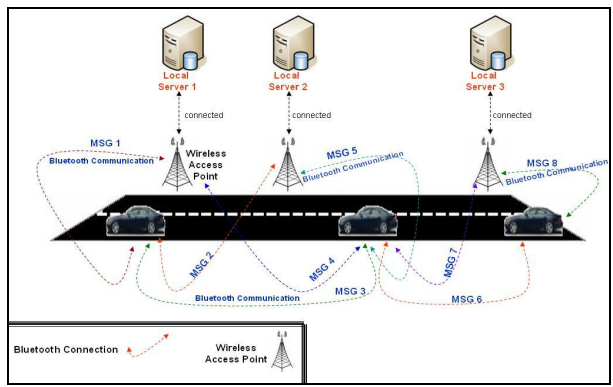

Figure 3. Messages are exchanged within the service infrastructure between (a) vehicle and base-station and (b) vehicle to vehicle using bluetooth connection.

*2) Alert Messages*

Consider the scenario in Figure 1. Vehicle M moving from the south to the north of King Street detects an accident on that road. Information about this accident condition (depicted by a red cross) should be propagated to other vehicles moving into that area. To this end, the vehicle that detects this condition generates an alert message and transmits it via the underlying service infrastructure. To support the transmission of traffic-alert information, a special message-type is transmitted. This message carries information such as a message type, a description of the alert, a unique identifier and the message expiration time. An alert message is communicated to all nearby base-stations and nearby passing vehicles; upon arrival to a base-station, the message is broadcast to all vehicles in that area. The vehicle that creates this alert does not expect any response.

Furthermore, the service infrastructure can easily support *persistent alert broadcasting* through a expiration time type attribute (<msg_expire>). A persistent alert arrives to a base-station is temporally stored in a temporal dataset (*Temporal Alert* repository) and broadcasted for periodic intervals for a pre-define amount of time based on the expiration time. The expiration time of the alert is set by the driver himself.

*3) Access Control*

Consider the scenario in Figure 1 again. The driver of vehicle X is moving to the west of William Street to reach the last building in that street where he actually works. While he approaches the building with his car, the wireless stations of his workplace along the roadside detect his presence and automatically alert him via the navigation system, running on a computer embedded in his vehicle or mobile phone, whether he wants the gate to be opened. He chooses "yes" and soon as he was only a few meters away from the gate, the latter opens automatically. This scenario clearly indicates that access to relevant resources in a mobile pervasive environment is not only based on the user's authenticity or domain of access but also largely influenced by the location of the user. An additional piece of information that this service could have considered before authorizing access to a resource is time. If it was around 2300 hours and the employee of vehicle X is not scheduled to work during night time, then the gate service alert would have never been sent to him.

To handle the above scenario and access control management in general, we have a flexible and dynamic access control component integrated within the service infrastructure that uses contextual information such as user's information (e.g., working schedule), location (e.g., near workplace) and time (e.g., 2300 hrs) information before deciding upon the allocation of a resource.

Positional data is very important for the internetworking between vehicles and is exchanged with time stamps. This is a critical parameter in any considered application due to the high mobility of all the vehicles. For example a vehicle may move in and out of networks range, but using positional information and speed it had acquired at a certain time and comparing it with current time it would be able to predict the vehicles location in its immediate vicinity and try to establish direct links with the relevant ones. Another method to deal with Ad hoc networking for inter-vehicle communication is by using the concept of clusters. In Bluetooth clusters are formed automatically when piconets are created. Here each vehicle or node belongs to a particular cluster having a unique ID, which would be the master's address. Each piconet (cluster) can be thought of as having a virtual perimeter around it. There are two types of perimeters which could be constructed around the devices. First, the master of each piconet would transfer information (tables as in the previous case) amongst its slaves. Secondly, the slaves falling on the edge of the perimeter would be made the bridging nodes (relay nodes), such that inter-piconet communication can take place. Each device would need to be aware of the neighbouring piconets, whose ID and perimeter values would be communicated through the bridging nodes. Knowledge of the neighbouring piconets would help in nodes migrating to other piconets to form quick connection since the masters ID would be known. All the above issues are subject to investigation.

## III. ARCHITECTURAL DESIGN

In this section, the service infrastructure is explored with detailed description of the different modules of the system. It consists of 2 main parts: the server component (resides on the local servers) and the client component (resides on the Bluetooth enabled mobile phone or embedded sensor vehicle). Figure 4 and 5 show the design architecture of the server and client components respectively. We also describe the message format used in the service infrastructure.

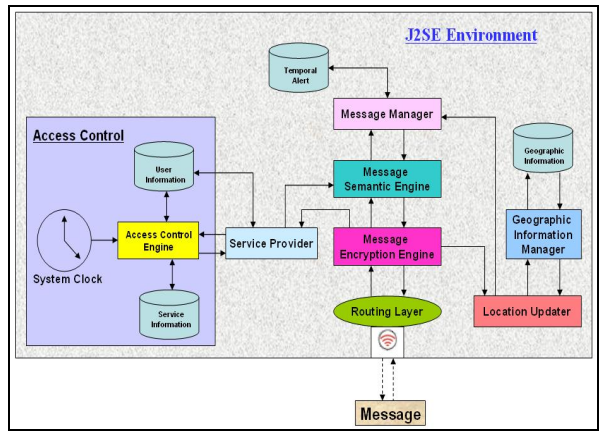

Figure 4. Server Component Architecture

The service infrastructure is based on an extended client-server computing model. In this model, a driver queries information via Bluetooth-enabled mobile device about traffic conditions or available facilities on a particular road segment. This query is sent toward that road segment, via the underlying ad-hoc network infrastructure. Vehicles in the destination area collaborate to establish a server, resolve the incoming queries and send back messages carrying the requested information. Both the client and server components specify the format and the semantics of the messages exchanged between clients (vehicles) and servers (base-stations).

The functions of the different modules of the server component architecture are as follows:

The **Geographic Information** repository contains a set of digitized map records representing the geographic area that is covered by the service infrastructure. Locations are represented as three-value tuples [road_id, road_name, segment_id], where *road_id* is a unique key representing a road, *road_name* is the name of the road and *segment_id* is a number representing a segment of that road; opposite traffic direction on the same road are represented as different road-segments. The **Geographic Information Manager (GIM)** is responsible for the update of geographic information. It also acts as an interface for retrieving geographic information. To this end, the GIM has to continuously response to request for geographic information from the Location Updater.

The **Location Updater** has the task of continuously detecting vehicles that are new to a particular road and/or road segment and consequently, send the latter a message to update its geographic location information. To help in its task, the Location Updater interacts with the GIM.

**Routing Layer** deals with the geographic routing protocol of the server. It is responsible for route discovery between the source and the target location areas using the service infrastructure. An intelligent broadcast with implicit acknowledgement to improve system performance by limiting the number of messages broadcast within the network for a given emergency event. If the event-detecting vehicle receives the same message from behind, it assumes that at least one vehicle in the back has received it, and stops broadcasting. The assumption is that the vehicle in the back will be responsible for moving the message along to the rest of the network. Note that it is possible for a vehicle to receive a message more than once, forwarded by different vehicles in the front. If this happens, the vehicle only acts on the first message.

The **Server Message Manager** (SMM) has the responsibility of managing incoming and outgoing messages based on the message type (msg_type) attribute of the latter. Incoming messages can be simply query messages that are forwarded, alert messages that are continuously forwarded for a pre-defined amount of time and response to a server query to access a service. Outgoing messages are normally generated by the server to be sent to a client/vehicle. For e.g., alerting the driver whether gate access to be opened. Before this message could be sent, it is generated by the service provider requesting the driver's identification.

The **Message Semantic Engine** (MSE) implements the service infrastructure Message Format Specification (MFS). It focuses on the format of messages. The MSE has the tasks of decomposing and composing the messages based on the MFS which is covered in the next section.

The **Message Encryption Engine** (MEE) encrypts and decrypts both incoming and outgoing messages based on an encryption and decryption algorithms that we have developed.

**Temporal Alert** is a repository stores temporarily alerts that have been scheduled to be broadcast for a pre-defined amount of time. Management of the Temporal Alert repository is performed by the MM.

The **Service Provider** module has to continuously sense vehicles that are new to a particular road and consequently, send the latter a message to request the driver's identification information. This information is represented using a two-value tuples [username, code]. It then check for user/driver authenticity, based on the latter's response, in the *user*

361

*information* repository to decide whether a service alert should be sent to him.

The **Access Control** (AC) module has the task of controlling access to the resources provided by the server. The proposed AC mechanism is simple but dynamic and flexible in the sense that the Access Control Engine (ACE) provides flexibility in the utilization and management of access control rules/policies embedded in the ACE. The Service Information repository contains the different services and the User Information dataset stores the users' profiles, preferences, and past interactions. Upon the receipt of a request made available by the Service Provider, the ACE also acquires the location of the service requestor. It then uses location information, time, user information and policies to decide whether to grant an access or not.

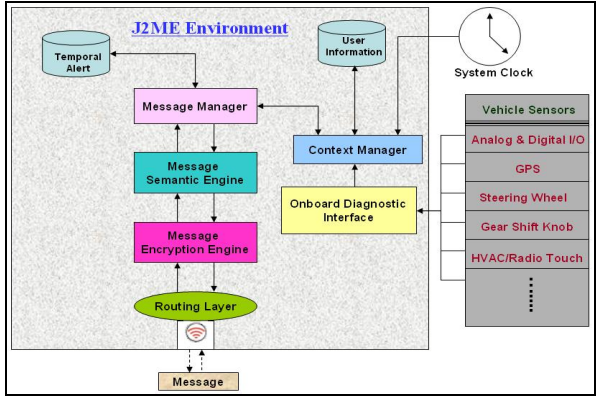

Figure 5.  Client component architecture

Modules in the client component architecture (Figure 5) such as Message Semantic Engine, Message Encryption Engine, Routing Layer, Temporal Alert repository are similar to those found at the server-side. Below are descriptions of the remaining modules:

The **Client Message Manager** (CMM) has the responsibility of managing incoming and outgoing messages based on the message type (msg_type) attribute of the latter. Incoming messages are messages that are sent by local servers and/or other vehicles. Examples are, query about traffic condition, alert message, request for identification by a server and service alerts. Outgoing messages are messages that are created to respond to incoming messages.

The **Onboard Diagnostic Interface** is responsible to read-in the different values captured by sensors embedded in the vehicle.

**User Information** repository holds information about the user profile, preferences, and behavior in accessing services.

The **Context Manager** provides the MM with context information, for e.g., speed, to determine rate of traffic flow. It also alert the driver of the different services available based on user, time and location information.

The structure of the message format that is exchanged in the service infrastructure is shown in Figure 6.

Figure 6.  Message Format

The TYPE attribute takes the values query, alert or service; where query represents a request that queries attributes of an area, alert represents the dissemination of alert messages and service represents a request for a service. The TARGET and SOURCE specifies the target and source-location of a message respectively. The ID and CREATOR indicate the message id and the vehicle id that generated this message respectively. The EXPIRES and COUNT attributes are used to end the dissemination of an alert or a query request. EXPIRES take an amount of time in seconds and this value is constantly checked against the TIME attribute which represent the time the message was created. COUNT stores the number of vehicles that can receive the message. Finally, the BODY represents the message body.

IV.  SIMULATIONS AND DISCUSSIONS

The main challenge in characterizing Bluetooth's performance over large distances is to determine the maximum range over which devices embedded in moving vehicles can discover one another and establish connections. This is a very simple test, but the results proved surprising and require significant consideration. The Class 1 (20dBm) Bluetooth device has a maximum range of 225 m compared to its specification range on 100 m. This result is surprising and the over-performance is a very encouraging indication of Bluetooth's feasibility for use in mobile environments.

TABLE I.  EFFECT OF SPEED ON CLASS 1 (20 DBM) BLUETOOTH DEVICES

| Vehicle Speed | | Max. Time "In Range" vehicle Bluetooth connection (sec) |
|---|---|---|
| km/h | m/s | |
| 100 | 27.8 | 7 |
| 80 | 22.2 | 10 |
| 60 | 16.7 | 13 |
| 40 | 11.1 | 18 |
| 20 | 5.6 | 35 |

Table 1 shows the speed calculations for Class 1 (20dBm) device using a nominal range of 100m according to Bluetooth specification. In these examples, one Bluetooth device is moving at a constant speed, and the other is stationary. The collaborative driving application involves outdoor transmissions occurring at high speeds, hence, a Class 1 Bluetooth device (100m range) will be appropriate. It is clear from the measurements in Table 1 that a lengthy connection setup time will severely inhibit Bluetooth's usefulness in short-term ad hoc connections between two rapidly moving vehicles.

The discovery time of different Bluetooth device was measured. The discovery time can theoretically vary between devices if the clocks are not initially well synchronized, as can be the case in devices from different manufacturers. Table 2 shows the results. We can see from this experiment that the average connection time to an individual Bluetooth device is, indeed, around 2s.

TABLE II. CONNECTION TIME TO 1 BLUETOOTH DEVICE

|  | Number of Tests | Mean Discovery Time (seconds) |
|---|---|---|
| **Vehicle Device 1** | 20 | 2.25 |
| **Vehicle Device 2** | 20 | 2.11 |
| **Vehicle Device 3** | 20 | 2.33 |
| **Vehicle Device 4** | 20 | 2.60 |

We are currently expanding our hardware testbed to include interferences and real obstacles for Bluetooth devices on high traffic moving vehicles. This expanded testbed will allow us to perform an even larger variety of connection setup, throughput and routing experiments with Bluetooth. The simulation results are also very promising in the ability, by making some very simple modifications to the Bluetooth baseband protocol; to reduce the time it takes for Bluetooth devices to discover other Bluetooth devices.

## V. CONCLUSIONS AND FUTURE WORKS

With the proliferation of wireless broadband technologies and the development of advanced smart phones, the transportation and logistics industry is well poised to leverage the advantages of ubiquitous connectivity to provide more reliable, efficient and responsive services. In this paper, we explored a mobile software architecture for bridging the gap between planning and execution in a pervasive transportation landscape empowered by next generation wireless communication and smart phone technologies. We identified key challenges in developing the mobile infrastructure and report on lessons learnt in implementing a prototype simulator for a generic Java 2 Mobile Edition (J2ME) enabled mobile device.

We also presented a service infrastructure designed to support collaborative driving in a mobile pervasive environment. From experiment results, we conclude that Bluetooth nodes Class 1 is acceptable for inter-vehicle communication within range. Nevertheless, we intend to study different vehicular and wireless networks simulation tools [9][10][15][17][18] and we plan to conduct simulation studies of large scale vehicular networks to evaluate our model in dense urban traffic for proven feasibility of our approach for collaborative driving in pervasive environments.